\documentclass[10pt,aps,prl,twocolumn,notitlepage,superscriptaddress,preprintnumbers]{revtex4-1}
\usepackage[utf8]{inputenc}
\usepackage[utf8]{inputenc} 
\usepackage[T1]{fontenc}
\usepackage{amsmath,amssymb,amsfonts}
\usepackage{bm}
\usepackage{graphicx,color}
\usepackage{verbatim}
\usepackage{float}
\usepackage[FIGTOPCAP]{subfigure} 
\usepackage{dcolumn}
\usepackage{natbib}
\usepackage{hyperref}
\usepackage{tikz}
\usetikzlibrary {arrows.meta} 
\usetikzlibrary{patterns}
\usetikzlibrary{hobby}
\usepackage{appendix}
\usepackage{ulem}

\newcommand{\RN}[1]{%
	\textup{\uppercase\expandafter{\romannumeral#1}}%
}

\begin{document}
	\title{Frustration of triplet interaction in spin-glass background}
	\author{M. Bagherikalhor}
	\email{mahsa.bagherikalhor@gmail.com}
	\affiliation{Department of Physics, Shahid Beheshti University, Evin, Tehran 1983969411, Iran}
	\author{B. Askari}
    \email{b.askari.sbu@gmail.com}
	\affiliation{Department of Physics, Shahid Beheshti University, Evin, Tehran 1983969411, Iran}
	\author{G.R. Jafari}
	\email{gjafari@gmail.com}
	\affiliation{Department of Physics, Shahid Beheshti University, Evin, Tehran 1983969411, Iran}
	\affiliation{Institute of Information Technology and Data Science, Irkutsk National Research Technical University, Lermontova St., 664074 Irkutsk, Russia}
	\affiliation{Center for Communications Technology, London Metropolitan University, London N7 8DB, UK}

	\date{\today}
	
\begin{abstract}
Parisi demonstrated in 1979 that pairwise interactions exhibit a glass spin phase when there is disorder. While he discovered an equilibrium solution of the Sherrington-Kirkpatrick (SK) spin-glass model and we know it as a continuous phase transition, the model dedicated to pairwise interactions and higher-order interactions has not been addressed.
This research intends to determine whether this phase exists in triplet interactions. Due to the intractable nature of the three interacting spins alone, we employed a perturbation approach to provide an analytical solution for the triplet interactions in the background of the SK spin-glass model. Our results show the existence of this phase in the third-order interaction and a sudden transition that indicates a change in the nature of a glassy spin system transitioning from the continuous order to the first order. It causes a forward shift in the critical temperature by identifying the frustration of triplet interactions. 
\end{abstract}

\maketitle
Over the years, great effort has gone into understanding the behavior of systems of spins interacting via quenched random couplings (spin-glasses). In $1975$ Sherrington and Kirkpatrick (SK) proposed an idealized model of a spin-glass~\cite{Sk1, Sherri} which is the infinite-range version of the Edwards–Anderson model~\cite{AndersonEdwards}. Methods for solving the SK model include generalization to models involving p-spin interactions~\cite{Kosterlitz,crisanti}. Derrida showed that under $p\rightarrow\infty$ the SK model is identical to a random energy model and is exactly solvable~\cite{Derrida1, Derrida2}. Thouless \textit{et. al} represented a solution to the SK model via the mean field equation so-called TAP equation~\cite{TAP}. Due to the frustration and disorder in couplings which are essential features of spin-glass, there are various local minimum states in the free energy landscape of this systems~\cite {Dotsenko, GrossMezard, crisanti}. In $1979-80$ Parisi proposed a solution with an interpretation of the structure of valleys of free energy~\cite{PARISI1979203, Parisi1980, GParisi1980, Parisi1, Parisi2} and the validity of Parisi's ansatz lies in its stability and its agreement with numerical experiments~\cite{Dominicis}. After the equilibrium solution found by Parisi for the SK spin-glass model and many years of efforts and focus on spin-glass~\cite{Thomas, Baity2013, Biroli, Baik2021, Newman}; there is still no analytical solution on the physical behavior of higher-order interacting systems.\\

The effects of higher-order interactions are summarized in Ref.~\cite{nature, Kuehn, Battiston, Majhi, bianconi_2021}. Studies show that going beyond the pairwise model by adding higher-order interactions can change the transition from continuous to discontinuous~\cite{nature, Kuehn}. Considering higher-order interactions and studying the collective behavior of the system by representing an analytical solution reports a discrete phase transition~\cite{Skardal, Battiston, Kargaran, Hakimi}. Research on triadic interactions confirms the occurrence of abrupt critical behavior in a system including three-body interactions~\cite{Huang, Arenas, mahsa, Zahedian}. It should be emphasized that three-body interactions can exist entirely independently of pairwise interactions and are not invariably the result of pairwise interactions. Although the pairwise interaction of spin pairs has been studied for years, the spin-glass of triplet interactions has not been reported yet. It is possible that it has not been accepted to be solved. Furthermore, the issue of the spin-glass phase in the triplet interactions, still has to be addressed, if it exists. In this letter, we investigate what will happen when triplet interactions are taken into account in addition to the background of paired spin interactions in the SK spin-glass model.\\

\begin{figure}[t]
	\hspace{-0.5cm}
	\includegraphics[scale = 1]{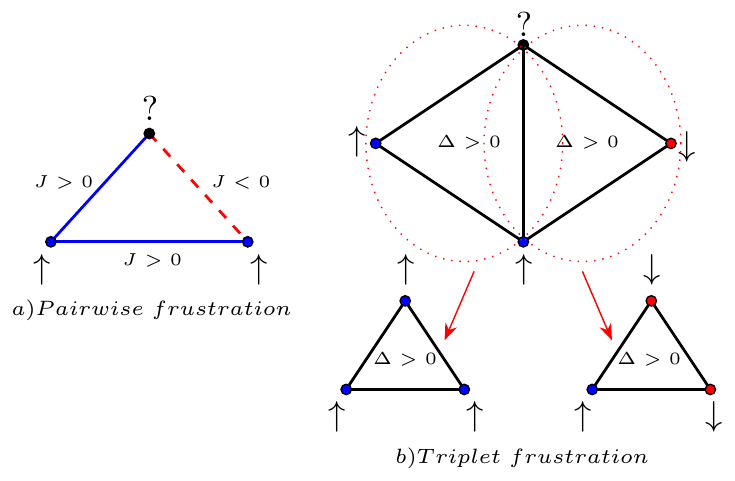}
	\caption{a) Spin frustration in a pairwise interacting spin-glass system. The frustrated node can not choose a state under the effect of a relationship with other nodes. Whether the frustrated node chooses an upward or downward direction the energy of the system does not change according to the Eq.\ref{equ1}, b) Schematically displays the frustrated state in triplet interaction when two triplets are placed together. Each triplet is not frustrated individually.}
    \label{fig:fig1}
\end{figure}

Frustration and the disorder on the pairwise interaction of spin pairs, $J_{ij}$, are thought to be the fundamental features in the SK spin-glass model~\cite{PWAnderson, doi:10.1142/9789812799371_0009}, and the most important consequence of frustration is that it leads to high degeneracy of the ground state of the system~\cite{Chowdhury}. By considering higher-order interactions we need to define the frustration for the groups of interacting spins. We introduce the frustration of triplet interactions when two triplets are placed together. According to the concept of frustrated state in the SK spin-glass model Fig.\ref{fig:fig1}.a, the frustrated node can not choose a state under the effect of a relationship with other nodes. In Fig.\ref{fig:fig1}.b, we schematically display triplet frustration where the black node in Fig.\ref{fig:fig1}.b chooses to be upward in the left triplet and downward in the right triplet while the juxtaposition of two triplets results in a frustrated state that the black node can not determine its direction.\\

Studying the phases of a complex system requires a parameter called the order parameter. In the normal ferromagnet Ising model, the magnetization is the order parameter which is zero for high temperatures and the system has only one state while it has two states of positive and negative in sufficiently low temperatures. For spin-glass, the situation is quite different. There are many equilibrium states~\cite{Sherri} and the order parameter would be sensitive to the existence of those which is a characteristic feature of glassy phase~\cite{Parisi1, Parisi2}. Following the Edwards-Anderson order parameter~\cite{AndersonEdwards}, the correlation between two spins is defined as an order parameter for the spin-glass which is called overlap~\cite{Castellani,nishimori}. \\

Considering the Hamiltonian for the SK spin-glass model with pairwise interactions,
{\footnotesize
\begin{equation}\label{equ1}
H_{J} = - \sum_{i<j} J_{ij} \ S_{i} \ S_{j} - h\sum_{i}S_{i},
\end{equation}
}
where the $J_{ij}$'s are independently random values taken from Gaussian probability distribution $\mathcal{P}(\mu_{_{J}},\,\sigma^2_{_{J}})$. The mean and the variance of pairwise interactions are proportional to $1/N$ for the reason that the energy needs to be extensive. To pursue our goal of considering higher-order interactions; we introduce the triplet interaction Hamiltonian as follows, 
{\footnotesize
\begin{equation}\label{equ2}
H_{\Delta} = - \sum_{i<j<k} \Delta_{ijk} \ S_{i} \ S_{j} \ S_{k},
\end{equation}}
the summation runs over all triplets of spins, the combination of $ N \choose 3 $, and each triplet is assigned by an independent random value, $\Delta_{ijk}$, from Gaussian probability distribution $\mathcal{P}(\mu_{_{\Delta}},\,\sigma^2_{_{\Delta}})$. The mean and the variance of this distribution should be proportional to $1/N^2$ to have Hamiltonian Eq.\ref{equ2} of order $N$. The primary goal of this study is to address the triplet spin interaction, which modifies the apparent nature of the transition. We look into how it affects the spin-glass behavior of the SK spin-glass model. The unsolvable nature of the triplet interaction of spins Eq.\ref{equ2}, led us to take advantage of a perturbation trick that solves the three interacting spins in the background of the SK spin-glass model. We aim to figure out what happens when we add triplet interaction as a perturbation term to the SK spin-glass Hamiltonian. Our proposed perturbation trick satisfies the following conditions
{\small
\begin{equation}\label{equ3}
\mu_{_{\Delta}} \ll \mu_{_{J}}, \quad \sigma^2_{_{\Delta}} \ll \sigma^2_{_{J}},
\end{equation}}
which means the mean and the variance of the triplet's random values are smaller than the pairwise ones by order of magnitude. To solve triplet interaction as a perturbation term in the background of the SK spin-glass model, we write the total Hamiltonian including the background and the perturbation term,
{\small
\begin{equation}\label{equ4}
H = H_{J} + H_{\Delta}.
\end{equation}}
Based on the proposed perturbation trick Eq.(3), the value of $H_{\Delta}$ is smaller than the value of $H_{J}$. Following, we represent the analytical solution for this system and our findings confirm the statement of~\cite{nature} that considering higher-order interactions leads to an abrupt transition. According to ~\cite{nishimori, mezard} 's approach for solving the SK spin-glass model with the replica method~\cite{Morone,vanHemmen}, we write the total partition function which includes the partition function for pairwise spin-glass and the triplet interactions.
\begin{widetext}
 We replicate the system $n$ times and calculate the configurational average of \textit{n}th power of the partition function then take the limit of $n\rightarrow0$. From the partition function, we manage to derive statistical quantities to describe the statistical properties of the system. The total partition function is $Z = Z_{_{J}} * Z_{_{\Delta}}$,
{\scriptsize
\begin{equation}\label{equ5}
[Z^n] = \int  \int  \left(\prod_{i<j} dJ_{ij} P(J_{ij})\right) \left(\prod_{i<j<k} d\Delta_{ijk} P(\Delta_{ijk})\right) Tr \exp \{ \beta \sum_{i<j} J_{ij} \sum_{\alpha=1}^{n} S_{i}^\alpha S_{j}^\alpha + \beta h \sum_{i=1}^{N} \sum_{\alpha=1}^{n} S_{i}^\alpha \} Tr \exp \{ \beta \sum_{i<j<k} \Delta_{ijk} \sum_{\alpha=1}^{n} S_{i}^\alpha S_{j}^\alpha S_{k}^\alpha \},
\end{equation}
}where $\alpha$ is the index variable for the replica. The integrals have been carried out and the result includes summations over $i<j$ and $\alpha, \beta$. We can rewrite these summation to $i$ , $\alpha<\beta$ in this way we have squared and cubic quantities of $\Bigl(\sum_{i} S_{i}^\alpha\Bigr)$, $\left(\sum_{i} S_{i}^\alpha S_{i}^\beta\right)$. To linearize those quantities by Gaussian integrals, we introduce the integration variables $q_{\alpha\beta}$ and $m_{\alpha}$ for the terms $\left(\sum_{i} S_{i}^\alpha S_{i}^\beta\right)$ and $\Bigl(\sum_{i} S_{i}^\alpha\Bigr)$,
{\scriptsize
\begin{equation}\label{equ6}
\begin{split}
&[Z^n] = \\
&\exp\left\{\frac{N n \beta^2 \sigma_{_{J}}^2}{4}\right\} \exp \left\{ \frac{N n \beta^2 \sigma_{_{\Delta}}^2}{12} - \frac{n \beta^2 \sigma_{_{\Delta}}^2}{4} - \frac{n \beta^2 \sigma_{_{\Delta}}^2}{12 N} \right\}\int \prod_{\alpha < \beta} dq_{\alpha\beta} \int \prod_{\alpha} dm_{\alpha}
\exp\left\{-\frac{N\beta^2 \sigma^2_{_{J}} }{2}\sum_{\alpha < \beta}q_{\alpha\beta}^2 - \frac{N\beta \mu_{_{J}}}{2}\sum_{\alpha}m_{\alpha}^2\right\} Tr\exp\Biggl\{\beta^2 \sigma^2_{_{J}}\sum_{\alpha < \beta}q_{\alpha\beta}\\
&\sum_{i} S_{i}^\alpha S_{i}^\beta + \beta \sum_{\alpha}(\mu_{_{J}} m_{\alpha}+h)\sum_{i}S_{i}^\alpha\Biggr\}
\exp \Biggl\{ \gamma \sum_{\alpha<\beta}q_{\alpha\beta}^2\sum_{i}S_{i}^\alpha S_{i}^\beta - (\frac{\beta^2 \sigma^2_{_{\Delta}}}{2N}+\frac{\beta^2\sigma^2_{_{\Delta}}}{6N^2})\sum_{\alpha<\beta}\sum_{i}S_{i}^\alpha S_{i}^\beta- (\frac{\beta \mu_{_{\Delta}}}{2N}+\frac{\beta \mu_{_{\Delta}}}{6N^2})\sum_{\alpha}\sum_{i}S_{i}^\alpha + \gamma'
\sum_{\alpha}m_{\alpha}^2 \sum_{i} S_{i}^{\alpha}\Biggr\}.
\end{split}
\end{equation}
}The $\gamma$, $\gamma'$ in the integrand are coefficients of terms $q^2_{\alpha\beta}$, $m^2_{\alpha}$. Calculating the integrals for $q_{\alpha\beta}$, $m_{\alpha}$ and applying our proposed perturbation trick results in the derivation of terms of $S_{i}^\alpha S_{i}^\beta$,  $S_{i}^\alpha$ to the power of three. This outcome is the consequence of considering triadic interactions. Now, by comparing the results of these integrals with one that has been derived from solving the integrals of the Eq.\ref{equ5}; the coefficients of the $\gamma$, $\gamma'$ would be derived. In the procedure of calculating the integrals of $dq_{\alpha\beta}$ and $dm_{\alpha}$, we have extended the denominators by the assumption that $\gamma$, $\gamma'$ are small values. 
\begin{equation}
{\footnotesize
\begin{aligned}\label{equ7}
&\int (\dots) \ dq_{\alpha\beta} = \exp\left\{\frac{(\beta^2 \sigma^2_{_{J}} \sum_{i}S_{i}^\alpha S_{i}^\beta)^2}{4(\frac{N\beta^2  \sigma^2_{_{J}}}{2}-\gamma\sum_{i}S_{i}^\alpha S_{i}^\beta)} - (\frac{\beta^2 \sigma^2_{_{\Delta}}}{2N}+\frac{\beta^2\sigma^2_{_{\Delta}}}{6N^2}) \sum_{\alpha < \beta} \sum_{i} S_{i}^\alpha S_{i}^\beta \right\} = \\
&\exp\left\{\frac{\beta^2 \sigma^2_{_{J}}}{2N}(\sum_{i}S_{i}^\alpha S_{i}^\beta)^2 + \frac{\gamma}{N^2}(\sum_{i}S_{i}^\alpha S_{i}^\beta)^3 - (\frac{\beta^2 \sigma^2_{_{\Delta}}}{2N}+\frac{\beta^2\sigma^2_{_{\Delta}}}{6N^2}) \sum_{\alpha<\beta}\sum_{i}S_{i}^\alpha S_{i}^\beta \right\}, \\
&\int (\dots) \ dm_{\alpha} = \exp\left\{\frac{\beta \mu_{_{J}}}{2N}(\sum_{i}S_{i}^\alpha)^2+\frac{\gamma'}{N^2}(\sum_{i}S_{i}^\alpha)^3 - (\frac{\beta \mu_{_{\Delta}}}{2N}+\frac{\beta \mu_{_{\Delta}}}{6N^2})\sum_{\alpha}\sum_{i}S_{i}^\alpha + \beta h \sum_{\alpha}\sum_{i}S_{i}^\alpha \right\}.
\end{aligned}
}
\end{equation}
\end{widetext}
 To find the coefficients of $\gamma$, $\gamma'$, we compare the partition function which is derived from Eq.\ref{equ6} by using the perturbation trick with ones from Eq.\ref{equ5} and we obtain that 
{\small
\begin{equation}\label{equ8}
\gamma = \frac{\beta^2 \sigma^2_{_{\Delta}}}{6} \quad , \quad\quad
\gamma' = \frac{\beta \mu_{_{\Delta}}}{6},
\end{equation}
}which perfectly agrees with our assumption that $\gamma$, $\gamma'$ are small values due to their dependence on $\mu_{_{\Delta}}$, $\sigma^2_{_{\Delta}}$. This is under our perturbation assumption Eq.\ref{equ3} that the mean value and the variance of the triadic interactions are smaller than those of pairwise interactions by order of magnitude. Therefore, our perturbation trick enables us to study the perturbation effect of triadic interactions in the background of the SK spin-glass model. In Eq.\ref{equ6} the terms proportional to $n$ and inverse of $N$ are ignorable by tending $n$ to zero and considering $N$ to be large. So, the coefficients of the terms $\sum_{\alpha<\beta}\sum_{i}S_{i}^\alpha S_{i}^\beta$, $\sum_{\alpha}\sum_{i}S_{i}^\alpha$ can be ignored in the thermodynamic limit. In Eq.\ref{equ6} the exponent of the integrand is proportional to $N$, so in the thermodynamic limit that $N\rightarrow \infty$ the integral can be evaluated by the steepest descent method. Also, we have represented the sum $\sum_{i}$ over a single site and considered the statement to the power of $N$.
\begin{equation}\label{equ9}
{\footnotesize
\begin{split}
&[Z^n]\simeq \exp\Biggl\{-\frac{N\beta^2 \sigma^2_{_{J}}}{2}\sum_{\alpha < \beta}q_{\alpha\beta}^2 - \frac{N\beta \mu_{_{J}}}{2}\sum_{\alpha}m_{\alpha}^2 + N log Tr e^{L'} \\
& + \frac{N n \beta^2 \sigma^2_{_{J}}}{4} + \frac{N n \beta^2 \sigma^2_{_{\Delta}}}{12}\Biggr\} \simeq  1+Nn\Biggl\{-\frac{\beta^2 \sigma^2_{_{J}}}{2n}\sum_{\alpha < \beta}q_{\alpha\beta}^2 \\
& - \frac{\beta \mu_{_{J}}}{2n}\sum_{\alpha}m_{\alpha}^2+ \frac{1}{n} log Tr e^{L'}+\frac{\beta^2 \sigma^2_{_{J}}}{4} + \frac{\beta^2 \sigma^2_{_{\Delta}}}{12}\Biggr\},
\end{split}
}
\end{equation}
In the last expression, the limit $n \rightarrow 0$ has been taken with $N$ kept very large but finite. Now, based on the replica method the free energy would be derived as, 
\begin{equation}\label{equ10}
{\footnotesize
\begin{split}
&-\beta[f] = \lim\limits_{n\rightarrow 0} \frac{[Z^n]-1}{nN} = \lim\limits_{n\rightarrow 0} \Biggl\{-\frac{\beta^2 \sigma^2_{_{J}}}{4n}\sum_{\alpha \neq \beta}q_{\alpha\beta}^2 - \frac{\beta \mu_{_{J}}}{2n}\sum_{\alpha}m_{\alpha}^2 \\
&+ \frac{1}{n}log Tr e^{L'} + \frac{\beta^2 \sigma^2_{_{J}} }{4} + \frac{\beta^2 \sigma^2_{_{\Delta}}}{12}\Biggr\}.
\end{split}
}
\end{equation}
where 
\begin{equation}\label{equ11}
{\footnotesize
\begin{split}
&L' =  \beta^2 \sigma^2_{_{J}} \sum_{\alpha < \beta}q_{\alpha\beta}S^\alpha S^\beta + \beta \sum_{\alpha}(\mu_{_{J}}m_{\alpha}+h)S^\alpha \\
&+ \gamma\sum_{\alpha<\beta}q_{\alpha\beta}^2 S^\alpha S^\beta + \gamma'\sum_{\alpha}m_{\alpha}^2 S^{\alpha},
\end{split}
}
\end{equation}
The values of $q_{\alpha\beta}$ and $m_{\alpha}$ should be chosen to extremize the quantity in the braces $\{ \}$ of Eq.\ref{equ10}. Hence, regarding the saddle-point condition, maximizing free energy Eq.\ref{equ10} with respect to $m_{\alpha}$, $q_{\alpha\beta}$ results in self-consistence equations,
\begin{equation}\label{equ12}
{\footnotesize
\begin{split}
& m_{\alpha} = \frac{Tr (S^\alpha+\frac{2\gamma'}{\beta \mu_{_{J}}}m_{\alpha}S^\alpha)e^{L'}}{Tr e^{L'}} \\
& M_{\alpha} \equiv \frac{Tr S^{\alpha}e^{L'}}{Tr e^{L'}} = \frac{m_{\alpha}}{1+\frac{2\gamma'}{\beta \mu_{_{J}}}m_{\alpha}} \\
& q_{\alpha\beta} = \frac{Tr(S^{\alpha}S^{\beta}+\frac{2\gamma}{\beta^2 \sigma^2_{_{J}}} q_{\alpha\beta}S^{\alpha}S^{\beta}) e^{L'}}{Tr e^{L'}} \\
& Q_{\alpha\beta} \equiv \frac{Tr S^{\alpha}S^{\beta} e^{L'}}{Tr e^{L'}} = \frac{q_{\alpha\beta}}{1+\frac{2\gamma}{\beta^2  \sigma^2_{_{J}}}q_{\alpha\beta}} 
\end{split}
}
\end{equation}
The variables $q_{\alpha\beta}$ and $m_{\alpha}$ that have been introduced as integration variables turn out to be related to the definition of order parameters, $Q_{\alpha\beta}$, $M_{\alpha}$, for our studied system while they represent the order parameters of the SK spin-glass model. The variable $Q_{\alpha\beta}$ represents overlap and $M_{\alpha}$ represents magnetization. These are normalized versions of order parameters and can be derived from $q_{\alpha\beta}$ and $m_{\alpha}$. These order parameters are explicitly dependent on replica indices. The replica method has been used for the convenience of calculating the configurational average. To derive the replica symmetric solution it has been assumed that $q_{\alpha\beta} = q$, $m_{\alpha} = m$ to discover that the replica indices should not affect the physics of the system. The symmetric solution for the free energy Eq.\ref{equ10} is
\begin{equation}\label{equ13}
{\footnotesize
\begin{split}
&-\beta[f] = -\frac{\beta^2 \sigma^2_{_{J}}}{4n}\{n(n-1)q^2\} - \frac{\beta \mu_{_{J}}}{2n}nm^2 + \frac{1}{n}logTr e^{L'} \\
&+ \frac{1}{4}\beta^2 \sigma^2_{_{J}} + \frac{1}{12}\beta^2 \sigma^2_{_{\Delta}},
\end{split}
}
\end{equation}
the third term including $L'$, in the right-hand side of the Eq.\ref{equ13}, can be calculated by using its definition from Eq.\ref{equ11} and a Gaussian integral. Inserting its result into Eq.\ref{equ13} and replaced the value of $\gamma$,$\gamma'$ from Eq.\ref{equ8} and taking the limit $n\rightarrow0$, we have the free energy as
\begin{equation}\label{equ14}
{\footnotesize
\begin{split}
&-\beta [f] = \frac{\beta^2 \sigma^2_{_{J}}}{4}(1-q)^2 + \frac{\beta^2 \sigma^2_{_{\Delta}}}{12}(1-q^2)-\frac{\beta \mu_{_{J}}}{2}m^2 \\
& + \int Dz \ log (2\cosh(\beta\tilde{H}(z))),
\end{split}
}
\end{equation}
where
\begin{equation}\label{equ15}
{\footnotesize
\begin{split}
&Dz = dz \exp(\frac{-z^2}{2}) \frac{1}{\sqrt{2\pi}} , \\
&\beta \tilde{H}(z) = \sqrt{\beta^2 \sigma^2_{_{J}} q + \gamma q^2}z +(\beta \mu_{_{J}} m + \beta h + \gamma' m^2).
\end{split}
}
\end{equation}
Extremizing the free energy concerning $m$, $q$, 
\begin{equation}\label{equ16}
{\footnotesize
\begin{split}
& m = (1+\frac{2\gamma'm}{\beta \mu_{_{J}}}) \int Dz \tanh(\beta \tilde{H}(z)) , \\
&q = \frac{1}{(1-\frac{\sigma^2_{_{\Delta}}}{3\sigma^2_{_{J}}})}\left\{1 - \frac{\beta^2 \sigma^2_{_{J}} + 2\gamma q}{\beta^2 \sigma^2_{_{J}}}\int Dz \frac{1}{\cosh^2(\beta\tilde{H}(z))}\right\}.
\end{split}
}
\end{equation}
These self-consistence equations need to be renormalized based on the definition of order parameters $M$, $Q$ Eq.\ref{equ12}. They satisfy our expectation that at the low temperature, the system goes to a ferromagnetic state where the magnetization and overlap equal one. In addition, in the limit of $\mu_{_{\Delta}}\rightarrow 0$, $\sigma_{_{\Delta}}\rightarrow 0$, equations for $M$ , $Q$ tend to their correspondences, $m$ , $q$ in the SK spin-glass concept. In the following, we draw the normalized order parameters $M$, $Q$, and $dQ/dT$ (derivative of overlap versus temperature) to illustrate how the system's behavior changes by considering triadic interactions. \\
\begin{figure}[ht] 
	\centering
    \includegraphics[scale = 0.32]{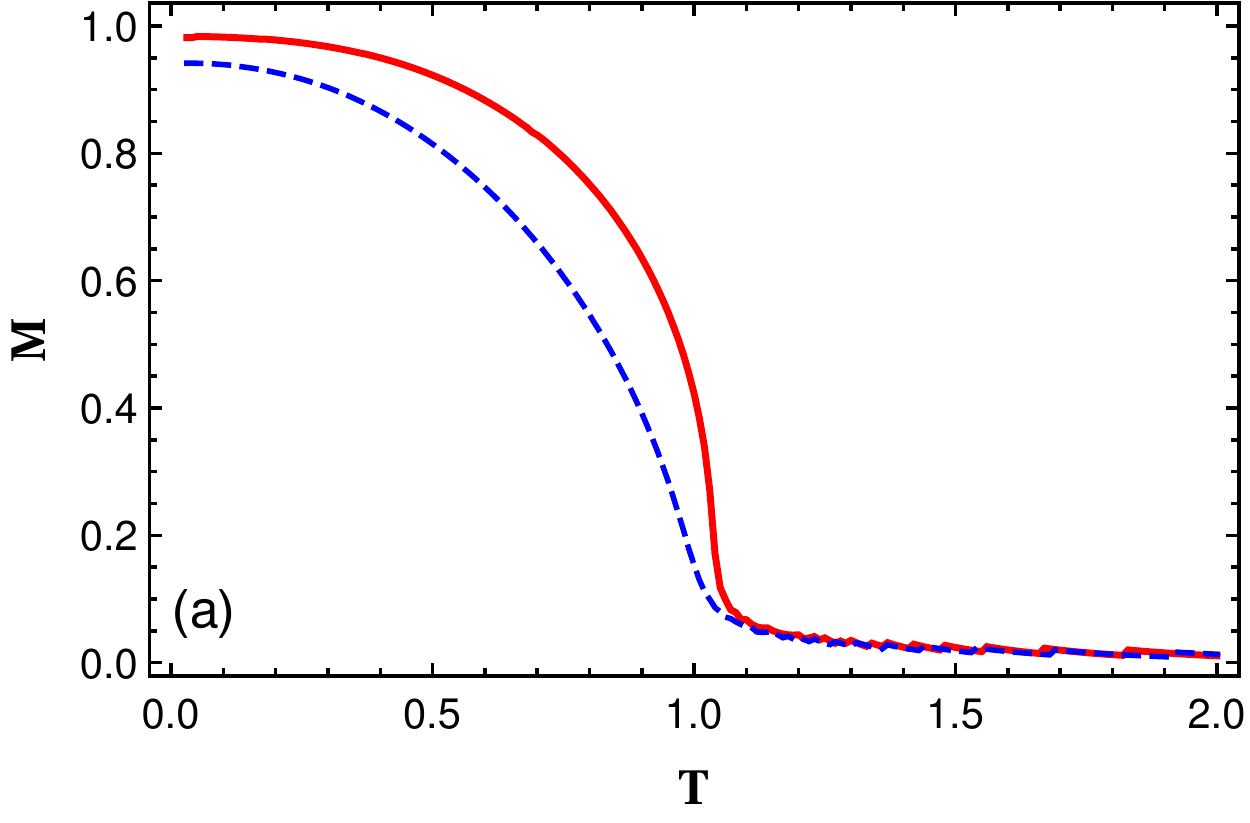}
    \vspace{0.2cm}
    \hspace{-0.2cm}
	\includegraphics[scale = 0.32]{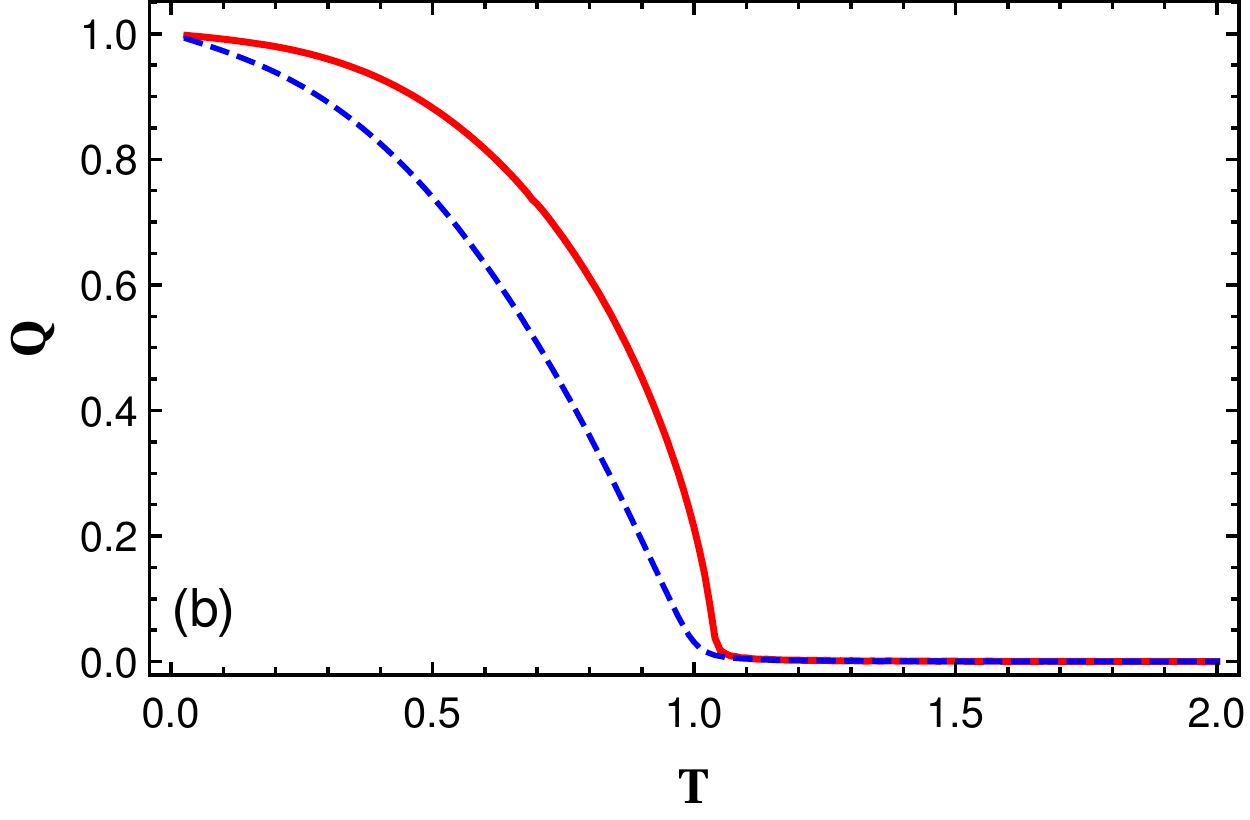}\\
    \hspace{-0.75cm}
	\includegraphics[scale = 0.69]{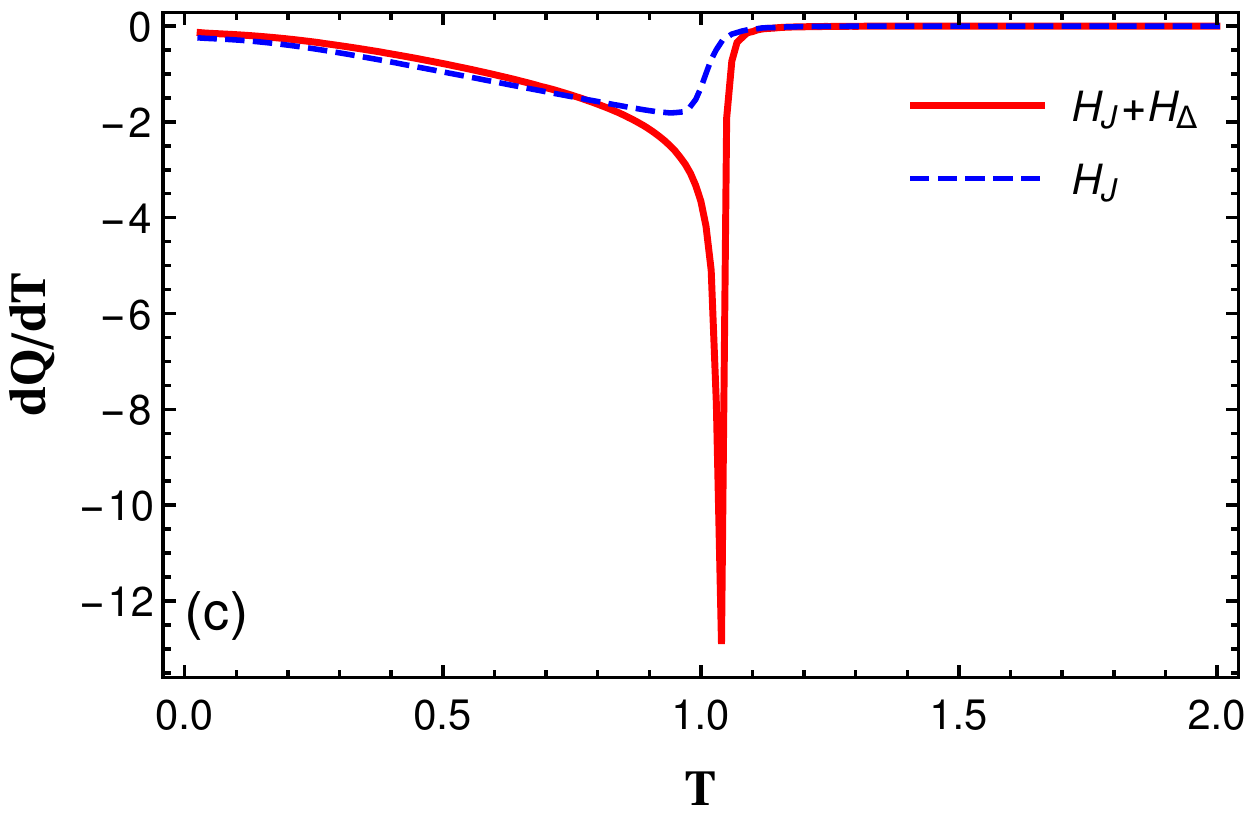}
	\caption{a) Magnetization as a function of temperature. Considering triadic interactions makes $M$ changes sharper. b) Overlap versus temperature is indicated. The critical temperature is altered when triadic interactions are taken into account. Also, an abrupt transition is seen by adding triadic interactions. c) Demonstrates the derivative of the overlap with respect to temperature. While it is constrained for the SK spin-glass, taking triadic interactions into account sharpens the derivation of the order parameter. 
    \label{fig:fig2}}
\end{figure}
Figure ~\ref{fig:fig2} shows the temperature dependence of the order parameters and $dQ/dT$ for pairwise interactions alone as well as in the presence of triplet interactions. While considering triplet interactions, we adjust the mean and variance of the Gaussian probabilities to follow our perturbation assumptions \ref{equ3}. Figure~\ref{fig:fig2}.a shows the temperature dependence of $M$ and its behavior under the effect of triplets. Continuous changes in the value of $M$ at each temperature change, occur with the steeper slope under consideration of triadic interactions. In figure ~\ref{fig:fig2}.b, we show the temperature dependence of the overlap. The figure shows a forward shift in the critical temperature and a sharp transition due to triplet interactions. In figure ~\ref{fig:fig2}.c, we show the overlap derivative with respect to temperature. While fig.~\ref{fig:fig2}.b emphasizes the abrupt transition is brought about by the inclusion of triplets fig.~\ref{fig:fig2}.c highlights this abrupt transition. Whereas the derivative of $dQ/dT$ is limited for the SK spin-glass, taking into account the triplet interactions sharpens the derivation of the overlap. Our discretization parameter determines the depth of derivation. The smaller the discretization, the deeper the derivation.\\

Scientists found that frustration and disorder are the two most crucial characteristics to have a spin-glass system. The two-spins with only nearest neighbor interaction similar to the Ising model was introduced by Edwards and Anderson. Then the long-range interacting pairs of spins was an exactly solvable model of a spin glass introduced by David Sherrington and Scott Kirkpatrick. Giorgio Parisi later discovered the model's equilibrium solution using the replica approach in 1979. Despite the fact only two-spin interactions were studied, an exact solution for p-spin interactions was done. When $p$ tends to infinity this model is known as a random energy model, however in the limited cases for $p > 2$ the problem is not solved yet.
Hence, it motivates us to address triplet interacting spins. The proposed questions that we find answers for include: I)how to generalize the frustration concept in the case of triplet interactions where two triangles are placed together, and II) whether the presence of frustration and disorder in triplet interactions can make a spin-glass system.
\begin{itemize}
  \item We introduced triplet frustration in a system including three interacting spins. Disorder is quenched random values which are assigned to triangles. Due to the availability of necessities, frustration and disorder, we expect to have a spin-glass system under considering higher-order interactions.
  \item The Hamiltonian with three interacting spins has not been solved, and no analytical solution exists yet. Nevertheless, we employed a perturbation trick to solve the three interacting spins in addition to the two-spin interactions in the SK spin-glass model. 
  \item We derive the magnetization and overlap as two order parameters that demonstrate the phase transition in the system. The temperature dependence of these two parameters indicates a forward shifting in the critical temperature. The system with three spin interactions make the transition at a higher temperature than the SK spin-glass model.
  \item Our findings indicate that the three-spin interactions play an important role in the system's dynamics. In compared to the SK spin-glass model paired interactions, the slope of the overlap as an order parameter is sharp due to triplet interactions. This means the type of transition is projected to change from second order to first order when the three-spins interactions increase. 
\end{itemize}

\bibliography{MyReferences}
\end{document}